%% file: main_acmart.tex
\documentclass[sigconf]{acmart}

\usepackage{booktabs} 
\usepackage{todonotes}
\usepackage{cleveref}
\usepackage{algorithm2e}
\usepackage{tabularx}
\usepackage{tikz}
\usepackage{multirow}

\usepackage{pgfplots}
\pgfplotsset{compat=1.14}

\usepackage{subcaption}
\setcopyright{rightsretained}

\acmDOI{}

\acmISBN{}

\acmConference[]{Submitted}{KDD}{2018} 
\acmYear{}
\copyrightyear{2018}

\acmPrice{}

\newcommand{\hornc}{\textsc{HornConcerto}\xspace}
\newcommand{\amie}{\textsc{Amie}\xspace}
\newcommand{\amieplus}{\textsc{Amie+}\xspace}
\newcommand{\sparql}{\textsc{Sparql}\xspace}

\begin{document}
\title{Beyond Markov Logic: Efficient Mining of Prediction Rules in Large Graphs}

\author{Tommaso Soru}
\orcid{0000-0002-1276-2366}
\affiliation{%
  \institution{AKSW, University of Leipzig}
  \city{Leipzig} 
  \state{Germany} 
  \postcode{04109}
}
\email{tsoru@informatik.uni-leipzig.de}

\author{Andr\'e Valdestilhas}
\affiliation{%
  \institution{AKSW, University of Leipzig}
  \city{Leipzig} 
  \state{Germany} 
  \postcode{04109}
}
\email{valdestilhas@informatik.uni-leipzig.de}

\author{Edgard Marx}
\affiliation{%
  \institution{Leipzig University of Applied Sciences}
  \city{Leipzig} 
  \state{Germany} 
  \postcode{04277}
}
\email{edgard.marx@htwk-leipzig.de}

\author{Axel-Cyrille Ngonga Ngomo}
\affiliation{%
  \institution{DICE, Paderborn University}
  \city{Paderborn} 
  \state{Germany} 
  \postcode{33098}
}
\email{axel.ngonga@upb.de}


\begin{abstract}
Graph representations of large knowledge bases may comprise billions of edges.
Usually built upon human-generated ontologies, several knowledge bases do not feature declared ontological rules and are far from being complete.
Current rule mining approaches rely on schemata or store the graph in-memory, which can be unfeasible for large graphs.
In this paper, we introduce \hornc, an algorithm to discover Horn clauses in large graphs without the need of a schema.
Using a standard fact-based confidence score, we can mine close Horn rules having an arbitrary body size.
We show that our method can outperform existing approaches in terms of runtime and memory consumption and mine high-quality rules for the link prediction task, achieving state-of-the-art results on a widely-used benchmark.
Moreover, we find that rules alone can perform inference significantly faster than embedding-based methods and achieve accuracies on link prediction comparable to resource-demanding approaches such as Markov Logic Networks.
\end{abstract}

%
%


\maketitle

\input{body}

\bibliographystyle{ACM-Reference-Format}
\bibliography{aksw,hornc,mandolin} 

\end{document}

%% file: body.tex
%

\section{Introduction}

The number of published triple-based datasets has been exponentially increasing in the last years.
Knowledge Bases such as DBpedia~\cite{dbpedia_jws_09}, UniProt~\cite{journals/nar/Consortium17a}, and LinkedGeoData~\cite{SLHA11} comprise more than one billion triples each.
While these knowledge bases span manifold domains, they are usually far from being complete.
Several approaches have been devised to tackle this problem, known as \textit{link prediction} or \textit{knowledge base completion}.
Most of these methods belong to the categories of translation-based approaches, tensor-factorization algorithms, neural networks, or general statistical-learning algorithms, some of which involving the use of first-order rules.
Beyond their utility in inference engines, rules help to understand insights in data, which is not a trivial task on large datasets.
In particular, rules have shown to be more explainable than factorization models, as they can be easily understood by humans as well as machines.

Rules have been specifically used in different environments, e.g. reasoners, rule-based systems, and Markov Logic Networks (MLNs).
Non-probabilistic reasoners utilize declared schemata describing a definition or a restriction, which imply -- for their nature -- a crisp reasoning.
For instance, with the restriction $\exists \mathtt{spouse} \sqsubseteq \mathtt{Person}$ (\textit{``All entities who have a spouse are persons.''}), one can infer that an entity is a person with a confidence value of $1.0$, if they have a spouse.
On the other hand, probabilistic reasoners assign a real confidence value between $0$ and $1$ to each statement, which can thus exist within some probability degree~\cite{wuthrich1995probabilistic}.
However, the growth of published structured data has made rule mining and probabilistic inference impracticable on certain datasets.
In order to optimize the computation, existing methods load data in-memory or rely on schemata while loading data into in-disk databases.
Unfortunately, classes and properties in ontologies are as complete as their instantiations.
For instance, in DBpedia 2016-04, among $63,764$ properties, only $3.5\%$ and $3.8\%$ of them are provided with a domain and range, respectively.\footnote{Retrieved on August 5th, 2017.}
Moreover, loading the largest available datasets in-memory is not feasible on a machine with average resources.

In this paper, we introduce an algorithm to the efficient discovery of Horn rules in graphs dubbed \hornc.
Using a standard fact-based confidence score, we can mine closed Horn rules having an arbitrary body size.
Our work brings the following contributions:
\begin{enumerate}
    \item We outperform existing rule mining approaches in terms of runtime and memory consumption, while scaling to very large knowledge bases.
    \item We mine high-quality rules for link prediction.
    \item We find that rules alone can perform inference significantly faster than embedding-based methods and achieve accuracies on link prediction comparable to resource-demanding approaches such as Markov Logic Networks.
\end{enumerate}

This paper is arranged as follows.
The next section presents the related work, while \Cref{sec:prel} introduces the preliminary concepts.
The \hornc approach is described in \Cref{sec:hornc}.
\Cref{sec:eval} shows the evaluation results, which are discussed in \Cref{sec:discussion}.
In \Cref{sec:conclusion}, we conclude.


\section{Related work} \label{sec:related}

This work is mainly related with Inductive Logic Programming (ILP) systems to mine rules in datasets where statements are composed by triples (or triplets).
The main differences among these approaches are in data storage type and the confidence score used for rules.
\textsc{Warmr} is a framework which combines ILP with association-rule mining, performing a breadth-first search to find frequent patterns. It uses a standard confidence function~\cite{dehaspe2001discovery}.
\textsc{Aleph} implements several measures for rule quality and has the peculiarity to generate random negative examples~\cite{muggleton1995inverse}.
With this respect, Sherlock, \amie, and \amieplus use only declared statements as counterexamples, thus embracing the open-world assumption (OWA)~\cite{schoenmackers2010learning,galarraga2015fast}.
While \amie uses an in-memory database to discover rules, Ontological Pathfinding (OP) -- state-of-the-art rule miner in terms of performances -- utilizes an external DBMS~\cite{chen2016ontological}.

Among the frameworks based on MLNs, the majority do not provide the possibility of discovering rules.
NetKit-SRL~\cite{macskassy2005netkit}, ProbCog~\cite{jain2010soft}, Tuffy~\cite{niu2011tuffy}, Markov theBeast~\cite{riedel08improving}, and RockIt~\cite{noessner2013rockit} all need rules as input in order to proceed with the weight learning, grounding, and inference phases. 
Alchemy~\cite{kok2009alchemy} instead performs the construction of rules in a combinatorial way; however, this method is highly inefficient, especially on datasets having an elevated number of distinct properties.
Besides link prediction and knowledge expansion~\cite{chen2014knowledge}, probabilistic inference systems showed to be effective in fields such as question answering~\cite{SHE+14}, instance matching~\cite{leitao2007structure}, and class expression learning~\cite{Buehmann2014}.


\section{Preliminaries} \label{sec:prel}

\subsection{Horn clauses}

In the context of first-order logic, an atom is a formula which contains a predicate and a list of undefined arguments and cannot be divided into sub-formulas.
For instance, the atom $A := a(x_1,x_2)$ features a binary predicate $a$ on the free variables $x_1, x_2$.
The typical structure of first-order rules expects an implication sign, an atom $H$ as the head (i.e., the implication target), and a set of atoms $B_1, \ldots, B_n$ as the body.
The body can be composed by one or more conjunction literals, i.e. $H \Longleftarrow B_1 \wedge \ldots \wedge B_n$.
An expression can be rewritten in disjunctive normal form (DNF) if it can be expressed as a chain of inclusive disjunctions.
Turning the rule above to the equivalent DNF, we have $H \vee \neg B_1 \vee \ldots \vee \neg B_n$.

\begin{definition}
\textbf{Horn clause.} \label{def:hc}
A Horn clause (or rule) is a disjunction of literals with at most one positive, i.e. non-negated, literal.
\end{definition}
Therefore, let $\overrightarrow{B} := \bigwedge_{i=1}^n{B_i}$, any first-order rule $H \Longleftarrow \overrightarrow{B}$ is also a Horn clause.

\subsection{Rule scoring}

Most rule mining algorithms measure and rank rules according to some score.
The most used measure for rule quality relies on the number of occurrences in data and is called \textit{standard confidence}.

\begin{definition}
\textbf{Standard confidence.} \label{def:stdconf}
The support of a Horn clause estimates the probability of the rule head to hold true, given the body:
\begin{equation}
    P(H|\overrightarrow{B}) = \frac{P(H \cap \overrightarrow{B})}{P(\overrightarrow{B})}
\end{equation}
The standard confidence of a rule $R := H \Leftarrow \overrightarrow{B}$ is defined as:
\begin{equation} \label{eqn:stdconf}
    c(R) = \frac{\left\lvert \{ H \wedge \overrightarrow{B} \} \right\rvert}{\left\lvert \{ \overrightarrow{B} \} \right\rvert} 
\end{equation}
The numerator in \Cref{eqn:stdconf} is also called support. 
For the sake of simplicity, we will name the denominator ``body support''.
\end{definition}

For instance, for $H := h(x,y)$ and $\overrightarrow{B} = B_1 := b_1(x,y)$, the standard confidence is a rate between two absolute numbers, i.e. (i) the number of occurrences found in graph $G$ such that $b_1(x,y) \wedge h(x,y)$ holds over (ii) the number of occurrences such that $b_1(x,y)$ holds.
We now formally define the concept of triple in a graph.

\begin{definition}
\textbf{Triple.} \label{def:rdf}
Given a directed labelled multi-graph $G=(V,E,l)$ with labelling function $l : V \cup E \rightarrow U$, where $U$ is the set of all possible labels, a triple $(s,p,o) := (\bar{s},\bar{p},\bar{o})$ belongs to the graph $G$ iff $s,o \in V \wedge e=(s,o) \in E \wedge l(s)=\bar{s} \wedge l(e)=\bar{p} \wedge l(o)=\bar{o}$.
\end{definition}

Algorithms such as \amie and Sherlock introduced a score which can deal with incomplete data~\cite{schoenmackers2010learning,galarraga2015fast}.
This score is referred to with the name \textit{partial completeness assumption (PCA) score}.

\begin{definition}
\textbf{PCA score.} \label{def:pca}
A Horn clause can be associated to a score called PCA score.
The rationale behind the PCA is that a missing relationship should not be treated as counterexample; instead, only triples $(s,p,o_1)$ can be treated as counterexamples for a triple $(s,p,o_2)$ with $o_1 \neq o_2$.
Being $\bar{H}$ any relationship $h(\cdot,\cdot)$, the PCA score of a rule $R := H \Leftarrow \overrightarrow{B}$ is defined as:
\begin{equation}
    pca(R) = \frac{\left\lvert \{ H \wedge \overrightarrow{B} \} \right\rvert}{\left\lvert \{ \bar{H} \wedge \overrightarrow{B} \} \right\rvert}
\end{equation}
\end{definition}


\section{Horn Concerto} \label{sec:hornc}

In this section, the core of the approach is presented.
\hornc is a rule mining algorithm designed with scalability in mind.
Moreover, we adopted a few constraints in order to let the computation focus on the more informative sections of the datasets first.
\hornc is a complete algorithm, meaning that if run without constraints, it can detect all Horn rules of a given type.
However, when dealing with large datasets, these constraints are necessary to reduce runtime and resource consumption.
The constraints are defined as such:
\begin{enumerate}
    \item consider only rules having a confidence value of at least $\theta \in [0,1]$;
    \item consider only rules having one of the top $P$ properties in its body.
    \item consider only top $T_i$ properties ($i \geq 2$) as part of a $|\overrightarrow{B}|=i$.
\end{enumerate}

\subsection{Algorithm}

The \hornc algorithm is shown in Algorithm~\ref{alg:hornc}.
First, we retrieve the top $P$ properties sorted by their occurrence counts.
The occurrence counts are the body support values for rules having a body size of 1.
Then, we mine the rules and compute their standard confidence (see Definition~\ref{def:stdconf}).
For each top-property $\bar{q}$, we count the occurrences of $p(x,y)$ and $\bar{q}(x,y)$ to happen together.
We can then find a map $m_1 : \Pi \rightarrow \mathbb{N}$ such that:
\begin{equation} \label{eqn:supp1}
  \bigcirc~~~m_1(p)=|\{(x,y) \in V^2 : (x,p,y),(x,\bar{q},y) \in G \wedge p \neq \bar{q} \}|\forall p \in \Pi
\end{equation}
where $\Pi=\{l(e) : e \in E\}$, i.e. the set of properties in the graph.
The confidence of a rule is thus the ratio between their support in \Cref{eqn:supp1} and body support.
Intuitively, changing $(x,y)$ into $(y,x)$ in the rule body, we can discover rules of type $p(x,y) \Leftarrow q(y,x)$, i.e. inverse properties.

\RestyleAlgo{boxruled}
\LinesNumbered
\begin{algorithm}
 \caption{\hornc} \label{alg:hornc}
 \KwData{target graph $G$, $\theta$, $P$, $T$}
 \KwResult{set $\mathcal{P}$ of mined rules and confidence values}
 retrieve the top $P$ properties\; 
 \For{each rule type \textbf{(in parallel)}}{
  initialize cache\;
  \For{each top property in $P$}{
   \eIf{rile body size $\leq$ $1$}{
        $\bigcirc$ Find matching digons and compute respective rule support\;
        Store matching digons into set $R$\;
   }{
     $\bigtriangleup$ Find matching triangles and compute respective rule support;\\
     Store matching triangles into set $R$\;
     \For{each rule in $R$}{
        $\fbox{}$ Compute body support (i.e., find \#adjacencies of body properties)\;
     }
   }
   \For{each rule in $R$ sorted by descend confidence}{
        \eIf{confidence $\geq$ $\theta$}{
            Add rule and its confidence to output set $\mathcal{P}$\;
        }{
            break\;
        } 
   }
  }
 }
\end{algorithm}


We then extend this concept to rules having a body size of 2.
These cases involving 3 nodes and 3 edges can be viewed as cliques of length 3 (or \textit{triangles}).
The aim is to compute a map $m^{(n)}_2 : \Pi^2 \rightarrow \mathbb{N}$ which returns the triangles involving properties $(p,r)$ given $\bar{q}$.

\begin{equation}
    \bigtriangleup~~~  m^{(n)}_2(p,r)=|\{(x,y,z) \in V^3 : (x,p,y),(z,r,y),(x,\bar{q},z) \in G\}| \forall p,r \in \Pi
\end{equation}


Consequently, rules of type $p(x,y) \Leftarrow \bar{q}(x,z) \wedge r(z,y)$ can be mined computing the absolute number of adjacencies among $\bar{q}$ and $r$; such values are the body support values for rules having a body size of 2.
\begin{equation}
    \fbox{}~~~  m^{(d)}_2(p,r)=|\{(x,y,z) \in V^3 : (x,\bar{q},z),(z,r,y) \in G\}| \forall p,r \in \Pi
\end{equation}

In the cases above, since free variables always appear at least twice in a rule, the rules are called \textit{closed}.
Closed rules have been preferred over open rules by existing algorithms, as they express propositions found in natural language~\cite{galarraga2015fast,chen2016ontological}.
Proceeding by induction, the maps above can be extended to mine closed rules having an arbitrary size.



In our evaluation, we mine six different types of Horn clauses, of which two are composed by two predicates (i.e., \textit{digons}) and four by three predicates (i.e., \textit{triangles}):

\begin{enumerate}
    \item $p(x,y) \Longleftarrow q(x,y)$
    \item $p(x,y) \Longleftarrow q(y,x)$
    \item $p(x,y) \Longleftarrow q(x,z) \wedge r(z,y)$
    \item $p(x,y) \Longleftarrow q(x,z) \wedge r(y,z)$
    \item $p(x,y) \Longleftarrow q(z,x) \wedge r(z,y)$
    \item $p(x,y) \Longleftarrow
    q(z,x) \wedge r(y,z)$
\end{enumerate}


\subsection{Implementation and complexity analysis}

\hornc was implemented
\footnote{The source code is available online at the address \url{https://github.com/mommi84/horn-concerto/}.}
 in Python 2.7 and the \sparql query language\footnote{https://www.w3.org/TR/sparql11-query/}, using Numpy and the Joblib library for multithread processing.
A Docker version of Virtuoso OpenSource 7.20 was used as triple store and query engine.
These architectural choices were motivated by the high availability of large knowledge bases in the so-called Web of Data~\cite{bizer2009linked}.
We therefore made \hornc compliant with the Resource Description Framework (RDF) and the \sparql query language.

The computational complexity of our implementation excluding constraints is linear with respect to the number of properties in the graph.
\cite{perez2009semantics} proved \sparql patterns to be \textsc{Pspace}-complete.
However, the time complexity relies on the \sparql engine which computes the queries.
The number of graph patterns in queries grows linearly with the size of the rule body.
Since our queries contain only \texttt{AND} and \texttt{FILTER} operators, they can be computed in a time $O(|\overrightarrow{B}| \cdot |E|)$, where $\overrightarrow{B}$ is the body size and $E$ is the total number of triples~\cite{perez2009semantics}.
We provided \hornc with optimizations based on caching, pruning, and parallel processing.
As queries for adjacencies might be repeated more than once, we introduced a caching system which stores the values and avoids unnecessary queries.
Moreover, as rules are sorted by confidence value in descending order, the algorithm can easily prune search spaces.
We implemented a parallel-processing version which executes each rule type on a different thread.


\section{Evaluation} \label{sec:eval}


In our setup, we evaluated the rule mining algorithms on four different measures: (1) execution runtime, (2) memory consumption (RAM and disk), (3) number of rules discovered, and (4) quality of rules for link prediction.

The datasets used are described in \Cref{tab:datasets}.
\texttt{D1-D2} are part of a benchmark for link prediction~\cite{TransE/bordes2013translating}.
Both datasets are divided into three parts, i.e. training, validation, and test set.
The sizes shown in the table are the union of the training and validation sets, while the test sets feature $5,000$ and $59,071$ triples respectively.
Datasets \texttt{D3-D4} are instead excerpts of the DBpedia knowledge graph~\cite{dbpedia_jws_09}.
\texttt{D3-D4}, unlike \texttt{D1-D2} which were manually annotated, are naturally incomplete and hence, the \textit{closed world assumption} does not apply.
We compared \hornc with two state-of-the-art rule mining approaches, i.e. OP and \amieplus.
We ran all three algorithms using their default settings; for \hornc, $\theta=0.001, P=200, T=10$. 
For a fair comparison in the runtime and memory evaluation, we set all approaches to utilize the same confidence thresholds. 
Finally, while \hornc and \amieplus expect only a triple dataset as input, OP expects an optional schema containing domains and ranges for each property.
We thus extracted the values of domain and range for every property featuring this information and added them to the OP input.
The experiments were carried out on a 8-core machine with 32 GB of RAM and running Ubuntu 16.04.

\setlength\tabcolsep{0.1cm}
\begin{table}[ht]
\centering
\caption{Datasets used in the evaluation.}
\begin{tabular}{llrrr}
\toprule
     & \textbf{Source} & \textbf{\# triples} & \textbf{\# nodes} & \textbf{\# prop.} \\
\midrule
\texttt{D1} & {WordNet (WN18)}     & 146,442     & 40,943    & 18 \\
\texttt{D2} & {Freebase (FB15k)}   & 533,144     & 14,951    & 1,345 \\
\texttt{D3} & {DBpedia 2.0 Person} & 7,035,092   & 2,299,075 & 10,341 \\
\texttt{D4} & {DBpedia EN 2016-04} & 397,831,457 & 5,174,547 & 63,764 \\
\bottomrule
\end{tabular}
\label{tab:datasets}
\end{table}


\begin{table*}[ht]
\centering
\caption{Results on all datasets. In the disk memory consumption, the dataset itself is not included.}
\begin{tabular}{clrrrr}
\toprule
\textbf{Dataset} & \textbf{Approach} & \textbf{Runtime} & \textbf{Rules} & \textbf{RAM (GB)} & \textbf{Disk (GB)} \\
\midrule
\multirow{ 2}{*}{\texttt{D1}}
& \amieplus & $20$ sec          & 151           & 0.1           & \textbf{$<$ 0.1} \\
& OP        & \multicolumn{4}{c}{\texttt{----- No Schema Available -----}} \\
& \hornc    & \textbf{12 sec}   & \textbf{365}  & 1             & 0.1 \\
\midrule
\multirow{ 2}{*}{\texttt{D2}}
& \amieplus & 2 h 45 min            & \textbf{45,427}   & \textbf{0.1}      & \textbf{$<$ 0.1} \\
& OP        & \multicolumn{4}{c}{\texttt{----- No Schema Available -----}} \\
& \hornc    & \textbf{2 h 01 min}   & 17,585            & 8                 & \textbf{1} \\
\midrule
\multirow{ 2}{*}{\texttt{D3}}
& \amieplus & $>5$ days & \textbf{$>$ 6,000} & 32 & \textbf{$<$ 0.1} \\
& OP & $>5$ days & N/A & \textbf{5} & $>1,000$ \\
& \hornc & \textbf{59 min} & 3,125 & 32 & \textbf{1} \\
\midrule
\multirow{ 2}{*}{\texttt{D4}}
& \amieplus & \multicolumn{4}{c}{\texttt{----- Out Of Memory -----}} \\
& OP & $>5$ days & N/A & \textbf{2} & $>1,000$ \\
& \hornc & \textbf{11 h 34 min} & \textbf{1,401} & 32 & \textbf{5.6} \\
\bottomrule 
\end{tabular}
\label{tab:compD3D4}
\end{table*}

\subsection{Rule mining results}

The results of the comparative evaluation on runtime, rules discovered, and memory consumption can be seen in~\Cref{tab:compD3D4}.
Only \hornc could complete the computation on both datasets, as \amieplus threw an out-of-memory error due to the large size of \texttt{D4} and OP started but did not finish to fill out its DBMS tables within 5 days.
\hornc discovered less rules (i.e., 1,401) on \texttt{D4}, which can be considered a superset of \texttt{D3}.
This is explained by the fact that the confidence score depends on certain ratios in data; increasing the dataset section does not necessarily lead to the same ratio values.

\begin{figure*}[htbp]
\centering
\begin{minipage}{0.45\textwidth}
  \centering
    \begin{tikzpicture}
    \begin{axis}[
    	scale=0.50,
    	legend pos=outer north east,
    	xlabel=number of CPUs,
    	ylabel=Time in minutes
    ]
    \addplot table [x=cpu, y=runtime, col sep=comma] {runtime.csv};
    \end{axis}
    \end{tikzpicture}
\subcaption[]{}\label{fig:parall}
\end{minipage}%
~
\begin{minipage}{0.45\textwidth}
  \centering
    \begin{tikzpicture}
    \begin{axis}[
    	scale=0.50,
    	legend pos=outer north east,
    	xlabel=Gibbs iter.,
    ]
    \addplot table [x=gibbs, y=hits1, col sep=comma] {mln.csv};
    \addplot table [x=gibbs, y=hits3, col sep=comma] {mln.csv};
    \addplot table [x=gibbs, y=hits10, col sep=comma] {mln.csv};
    \legend{
			hits1,
			hits3,
			hits10,}
    \end{axis}
    \end{tikzpicture}
\subcaption[]{}\label{fig:mln}
\end{minipage}%

    \label{fig:parmln}
    \caption{Effects of parallelization on runtime (left) and Gibbs sampling iterations on accuracy (right) on dataset \texttt{D3}.}
\end{figure*}
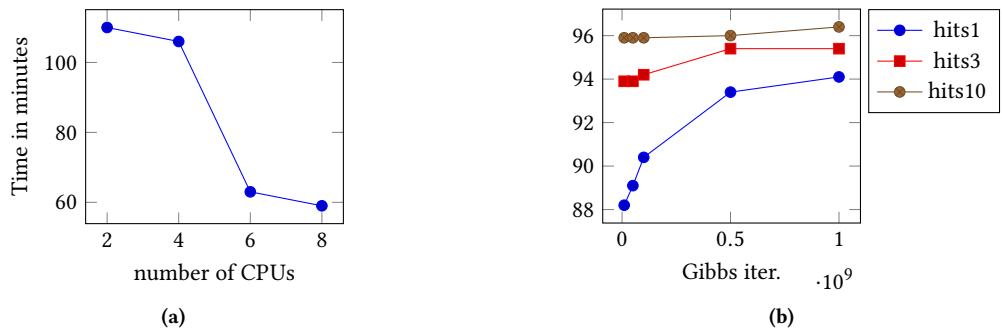


To assess the scalability of our method, we executed it using different numbers of CPUs.
We evaluated how parallelism and the use of CPUs affect the runtime of the rule mining process.
\Cref{fig:parall} shows that the runtime remains almost the same until 4 CPUs, drops at around 6, and then stabilizes again after 6 CPUs.
The reason is because we parallelize the computation on rule types, which are 6 in total.


\subsection{Link prediction results}

We compared \hornc with \amieplus and other state-of-the-art approaches on the link prediction task using datasets \texttt{D1} and \texttt{D2}.
\texttt{D3} and \texttt{D4} are not suited for link prediction, since there the \textit{closed world assumption} does not apply, meaning that a missing edge is not necessarily false.
As previously mentioned, the OP algorithm relies on schemata to build rules from data.
Being \texttt{D1-D2} both schema-free datasets, all executions of OP returned $0$ rules discovered, forcing us to abort its evaluation on link prediction.
Then, we compared their performance with the aforementioned \textsc{TransE}~\cite{TransE/bordes2013translating} and the following: 
\textsc{HolE}, a state-of-the-art algorithm for link prediction based on holographic embeddings and neural networks by~\cite{nickel2015holographic};
\textsc{ComplEx}, a latent-factorization method by~\cite{trouillon2016complex};
\textsc{Analogy} by~\cite{liu2017analogical}, an approach to link prediction using analogies among nodes embedded in a vector space;
an ensemble learning approach dubbed \textsc{DistMult}, by~\cite{kadlec2017knowledge}.

\begin{table*}[ht]
\centering
\caption{Link prediction results on \texttt{D1} and \texttt{D2}. Hits@N values are in \%. For each measure, the best result is highlighted in bold.}
\begin{tabular}{lcccccccc}
\toprule
 & \multicolumn{4}{c}{\texttt{D1} (WN18)} & \multicolumn{4}{c}{\texttt{D2} (FB15k)} \\
 & MRR & Hits@1 & Hits@3 & Hits@10 & MRR & Hits@1 & Hits@3 & Hits@10 \\
\midrule
\textsc{TransE} & 0.495 & 11.3 & 88.8 & 94.3 & 0.463 & 29.7 & 57.8 & 74.9 \\
\textsc{HolE} & 0.938 & 93.0 & 94.5 & 94.9 & 0.524 & 40.2 & 61.3 & 73.9 \\
\textsc{ComplEx} & 0.941 & 93.6 & 94.5 & 94.7 & 0.692 & 59.9 & 75.9 & 84.0 \\
\textsc{Analogy} & 0.942 & 93.9 & 94.4 & 94.7 & 0.725 & 64.6 & 78.5 & 85.4 \\
\textsc{DistMult E} & 0.790 & 78.4 & N/A & 95.0 & \textbf{0.837} & \textbf{79.7} & N/A & \textbf{90.4} \\
\midrule
\amieplus+Avg & 0.961 & 95.9 & 96.2 & 96.5 & 0.333 & 29.8 & 33.4 & 39.5 \\
\amieplus+Max & 0.970 & 96.8 & 97.1 & 97.1 & 0.352 & 32.2 & 35.3 & 39.8 \\
\amieplus+Prod & 0.927 & 91.9 & 92.9 & 94.6 & 0.336 & 30.6 & 33.7 & 38.7 \\
\amieplus+MLNs & 0.892 & 89.2 & 94.3 & 96.0 & 0.307 & 30.7 & 36.7 & 39.8 \\
\midrule
\hornc+Avg & 0.963 & 95.9 & 96.6 & 97.0 & 0.479 & 41.2 & 50.6 & 60.3 \\
\hornc+Max & \textbf{0.971} & \textbf{96.9} & \textbf{97.3} & \textbf{97.4} & 0.810 & 79.2 & \textbf{81.9} & 83.6 \\
\hornc+Prod & 0.941 & 91.8 & 96.4 & 97.2 & 0.508 & 44.7 & 52.7 & 61.9 \\
\hornc+MLNs & 0.904 & 90.4 & 94.2 & 95.9 & 0.224 & 22.4 & 31.4 & 37.2 \\
\bottomrule
\end{tabular}
\label{tab:linkprediction}
\end{table*}

In order to evaluate the rules discovered by \hornc and \amieplus on link prediction, we used four different settings.
The first three settings are based on the intuition which states that the probability of a missing link is proportional to the confidence scores of the rules involving its predicate.
We chose three different functions to compute this probability, i.e. average (\texttt{Avg}), maximum (\texttt{Max}), and product of complements (\texttt{Prod}).
These functions take the vector of the confidence scores as input.
The fourth setting relies on MLNs, instead.
We thus plugged the input datasets and the rules discovered by \hornc and \amieplus to \textsc{Mandolin}\footnote{https://github.com/AKSW/Mandolin}, a framework based on MLNs.

Dataset \texttt{D1-D2} are split into training, validation, and test set.
For each triple in the test set, corrupted triples are generated by altering subjects and objects.
Then, the probability of the triple is computed using the functions described above and MLNs.
The \textit{mean reciprocal rank} is obtained by inverting the average rank (by descending probability) of a correct triple among the corrupted ones.
Hits@$N$ measures the percentage of cases where a correct triple ranks in the top-$N$ triples.
Link prediction results are shown in~\Cref{tab:linkprediction}.
Results show that \hornc achieves the best Hits@3 on \texttt{D2} and state-of-the-art accuracies on dataset \texttt{D1}, where the differences among the two rule mining approaches are relatively small.


\begin{figure*}[htbp]
\centering
\begin{minipage}{0.45\textwidth}
  \centering
    \includegraphics[width=\textwidth]{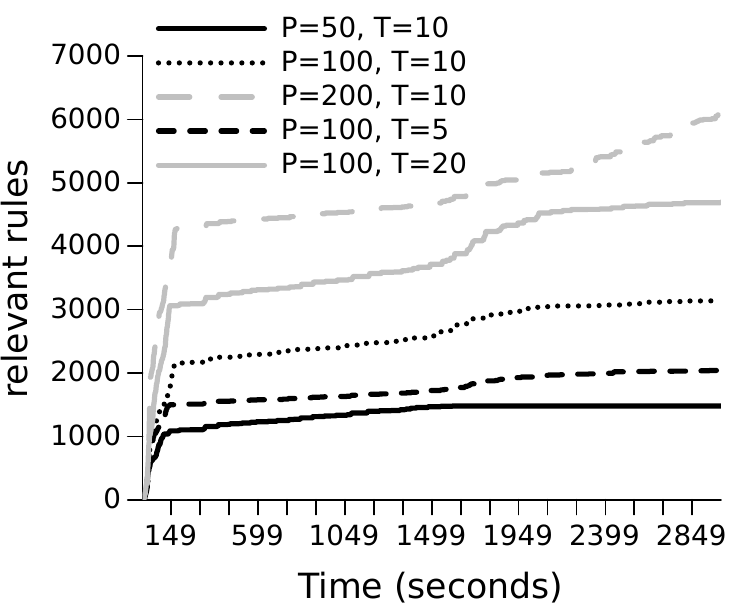}
\subcaption[]{Relevant rules}\label{fig:discovery}
\end{minipage}%
~~
\begin{minipage}{0.45\textwidth}
  \centering
     \includegraphics[width=\textwidth]{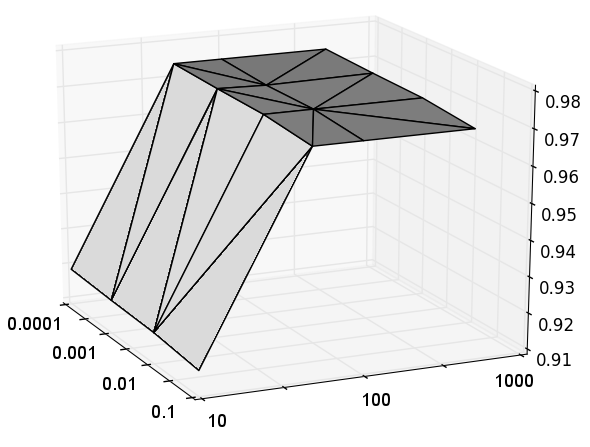}
\subcaption[]{Hits@10 accuracy}\label{fig:hitsat10}
\end{minipage}%
    \caption{}
\end{figure*}

In~\Cref{fig:discovery}, we plotted the number of relevant rules (i.e., rules meeting the confidence threshold $\theta=0.001$) discovered over time on dataset \texttt{D3}.
We launched 5 different settings, varying the number of top properties ($P=50,100,200$) and the number of adjacencies ($T=5,10,20$).
The configuration $\langle P,T \rangle$ which yielded the maximum number of rules is $\langle 200,10 \rangle$, while $\langle 50,10 \rangle$ produced less than one third of rules of the former.
Increasing $P$ or $T$ trivially leads to more rules discovered.
As can be seen, all the curves have a sudden braking at around $t=200s$, which is a consequence of having smaller (and thus simpler) rules be mined first.

In order to understand the effects of hyperparameters $\theta$, $P$, and $T$ on the accuracy, we ran \hornc on \texttt{D1} in a grid pattern.
We computed the Hits@10 accuracy on \texttt{D1}, varying $\theta=\{0.0001,...,0.1\}$ and $P=\{10,...,500\}$.
\Cref{fig:hitsat10} shows that choosing a smaller $\theta$ of one order of magnitude has little to no influence on the Hits@10 value; the same applies to $T$, which was set to $100$ in figure.
On the other hand, experiments with $P \geq 50$ achieved a score increment of +6\% w.r.t. $P=10$.


\section{Discussion} \label{sec:discussion}


As pointed out in \cite{galarraga2013amie}, implementing the PCA score in \sparql is inefficient.
In fact, we showed that the choice of using the PCA score instead of the standard confidence, even in systems not based on \sparql query engines, can lead to scalability issues.
Moreover, our evaluation suggests that approaches which rely on schemata as the only mean to achieve a good optimization still struggle with large datasets.
The concept of having to provide a property domain-range schema falls into contradiction with the need for rules, since this environment would more likely occur when no or few ontological insights are available.

The link prediction results we obtained were the most interesting ones.
Firstly, \hornc was able to perform better than \amieplus, in contrast with previous research where PCA was showed being more effective than standard confidence on the extraction of higher-quality rules.
More interestingly, we achieved the highest accuracy on Hits@N and MRR on \texttt{D1} and Hits@3 on \texttt{D2} just by using a simple maximization of the rule confidence scores.
The reason why dataset \texttt{D2} was harder to learn might be that rules having a longer body are needed to describe more complex relationships.
The fact that embedding-based approaches perform well could confirm this hypothesis.
However, these approaches are extremely slow at generating links; as their models compute a probability value for a given triple, yielding new links has a complexity of $O(|V|^2 \cdot |\Pi|)$.
Finally, we surprisingly found that, independently on the rule mining algorithm, rules alone can achieve higher accuracies on link prediction than cumbersome and resource-demanding approaches such as Markov Logic Networks.
Even after one billion Gibbs sampling iterations (see \Cref{fig:mln}), the links predicted through MLNs were not as good as the ones found by our three simple functions.
This can pave the way to an investigation on the existence of boundaries and theoretical limits that a rule-based prediction algorithm cannot overcome, due to mathematical constraints.


\section{Conclusion} \label{sec:conclusion}

In this paper, we presented \hornc, an algorithm for mining rules in large directed labelled graphs.
Our \sparql-based implementation has shown an unprecedented ability to scale on large datasets.
In the rule mining task, \hornc achieved state-of-the-art performances in execution runtime and disk memory consumption.
While these accomplishments were fulfilled, our algorithm managed to mine high-quality rules in datasets having hundreds of millions of triples without the need of a schema and achieve state-of-the-art link predictions on a widely-used benchmark.
Our findings suggest that (i) rule-based algorithms are still competitive on link prediction and (ii) rules alone can achieve accuracies comparable to cumbersome and resource-demanding approaches such as Markov Logic Networks.





%% file: main_acmart.bbl

\begin{thebibliography}{27}


\ifx \showCODEN    \undefined \def \showCODEN     #1{\unskip}     \fi
\ifx \showDOI      \undefined \def \showDOI       #1{#1}\fi
\ifx \showISBNx    \undefined \def \showISBNx     #1{\unskip}     \fi
\ifx \showISBNxiii \undefined \def \showISBNxiii  #1{\unskip}     \fi
\ifx \showISSN     \undefined \def \showISSN      #1{\unskip}     \fi
\ifx \showLCCN     \undefined \def \showLCCN      #1{\unskip}     \fi
\ifx \shownote     \undefined \def \shownote      #1{#1}          \fi
\ifx \showarticletitle \undefined \def \showarticletitle #1{#1}   \fi
\ifx \showURL      \undefined \def \showURL       {\relax}        \fi
\providecommand\bibfield[2]{#2}
\providecommand\bibinfo[2]{#2}
\providecommand\natexlab[1]{#1}
\providecommand\showeprint[2][]{arXiv:#2}

\bibitem[\protect\citeauthoryear{Bizer, Heath, and Berners-Lee}{Bizer
  et~al\mbox{.}}{2009}]%
        {bizer2009linked}
\bibfield{author}{\bibinfo{person}{Christian Bizer}, \bibinfo{person}{Tom
  Heath}, {and} \bibinfo{person}{Tim Berners-Lee}.}
  \bibinfo{year}{2009}\natexlab{}.
\newblock \showarticletitle{Linked data-the story so far}.
\newblock \bibinfo{journal}{\emph{Semantic services, interoperability and web
  applications: emerging concepts}} (\bibinfo{year}{2009}).
\newblock


\bibitem[\protect\citeauthoryear{Bordes, Usunier, Garcia-Duran, Weston, and
  Yakhnenko}{Bordes et~al\mbox{.}}{2013}]%
        {TransE/bordes2013translating}
\bibfield{author}{\bibinfo{person}{Antoine Bordes}, \bibinfo{person}{Nicolas
  Usunier}, \bibinfo{person}{Alberto Garcia-Duran}, \bibinfo{person}{Jason
  Weston}, {and} \bibinfo{person}{Oksana Yakhnenko}.}
  \bibinfo{year}{2013}\natexlab{}.
\newblock \showarticletitle{Translating embeddings for modeling
  multi-relational data}. In \bibinfo{booktitle}{\emph{Advances in Neural
  Information Processing Systems}}.
\newblock


\bibitem[\protect\citeauthoryear{B{\"u}hmann, Fleischhacker, Lehmann, Melo, and
  V{\"o}lker}{B{\"u}hmann et~al\mbox{.}}{2014}]%
        {Buehmann2014}
\bibfield{author}{\bibinfo{person}{Lorenz B{\"u}hmann}, \bibinfo{person}{Daniel
  Fleischhacker}, \bibinfo{person}{Jens Lehmann}, \bibinfo{person}{Andre Melo},
  {and} \bibinfo{person}{Johanna V{\"o}lker}.} \bibinfo{year}{2014}\natexlab{}.
\newblock \showarticletitle{Inductive Lexical Learning of Class Expressions}.
  In \bibinfo{booktitle}{\emph{Knowledge Engineering and Knowledge
  Management}}.
\newblock
\showISBNx{978-3-319-13703-2}
\urldef\tempurl%
\url{https://doi.org/10.1007/978-3-319-13704-9_4}
\showDOI{\tempurl}


\bibitem[\protect\citeauthoryear{Chen, Goldberg, Wang, and Johri}{Chen
  et~al\mbox{.}}{2016}]%
        {chen2016ontological}
\bibfield{author}{\bibinfo{person}{Yang Chen}, \bibinfo{person}{Sean Goldberg},
  \bibinfo{person}{Daisy~Zhe Wang}, {and} \bibinfo{person}{Soumitra~Siddharth
  Johri}.} \bibinfo{year}{2016}\natexlab{}.
\newblock \showarticletitle{Ontological Pathfinding}. In
  \bibinfo{booktitle}{\emph{Proceedings of the 2016 International Conference on
  Management of Data}}. ACM, \bibinfo{pages}{835--846}.
\newblock


\bibitem[\protect\citeauthoryear{Chen and Wang}{Chen and Wang}{2014}]%
        {chen2014knowledge}
\bibfield{author}{\bibinfo{person}{Yang Chen} {and} \bibinfo{person}{Daisy~Zhe
  Wang}.} \bibinfo{year}{2014}\natexlab{}.
\newblock \showarticletitle{Knowledge expansion over probabilistic knowledge
  bases}. In \bibinfo{booktitle}{\emph{Proceedings of the 2014 ACM SIGMOD
  international conference on Management of data}}. ACM.
\newblock


\bibitem[\protect\citeauthoryear{Consortium}{Consortium}{2017}]%
        {journals/nar/Consortium17a}
\bibfield{author}{\bibinfo{person}{The~UniProt Consortium}.}
  \bibinfo{year}{2017}\natexlab{}.
\newblock \showarticletitle{UniProt: the universal protein knowledgebase.}
\newblock \bibinfo{journal}{\emph{Nucleic Acids Research}}
  \bibinfo{volume}{45}, \bibinfo{number}{Database-Issue}
  (\bibinfo{year}{2017}), \bibinfo{pages}{D158--D169}.
\newblock
\urldef\tempurl%
\url{http://dblp.uni-trier.de/db/journals/nar/nar45.html#Consortium17a}
\showURL{%
\tempurl}


\bibitem[\protect\citeauthoryear{Dehaspe and Toivonen}{Dehaspe and
  Toivonen}{2001}]%
        {dehaspe2001discovery}
\bibfield{author}{\bibinfo{person}{Luc Dehaspe} {and} \bibinfo{person}{Hannu
  Toivonen}.} \bibinfo{year}{2001}\natexlab{}.
\newblock \showarticletitle{Discovery of relational association rules}.
\newblock In \bibinfo{booktitle}{\emph{Relational data mining}}.
  \bibinfo{publisher}{Springer}, \bibinfo{pages}{189--212}.
\newblock


\bibitem[\protect\citeauthoryear{Gal{\'a}rraga, Teflioudi, Hose, and
  Suchanek}{Gal{\'a}rraga et~al\mbox{.}}{2015}]%
        {galarraga2015fast}
\bibfield{author}{\bibinfo{person}{Luis Gal{\'a}rraga},
  \bibinfo{person}{Christina Teflioudi}, \bibinfo{person}{Katja Hose}, {and}
  \bibinfo{person}{Fabian~M Suchanek}.} \bibinfo{year}{2015}\natexlab{}.
\newblock \showarticletitle{Fast rule mining in ontological knowledge bases
  with AMIE+}.
\newblock \bibinfo{journal}{\emph{The VLDB Journal}} \bibinfo{volume}{24},
  \bibinfo{number}{6} (\bibinfo{year}{2015}).
\newblock


\bibitem[\protect\citeauthoryear{Gal{\'a}rraga, Teflioudi, Hose, and
  Suchanek}{Gal{\'a}rraga et~al\mbox{.}}{2013}]%
        {galarraga2013amie}
\bibfield{author}{\bibinfo{person}{Luis~Antonio Gal{\'a}rraga},
  \bibinfo{person}{Christina Teflioudi}, \bibinfo{person}{Katja Hose}, {and}
  \bibinfo{person}{Fabian Suchanek}.} \bibinfo{year}{2013}\natexlab{}.
\newblock \showarticletitle{AMIE: association rule mining under incomplete
  evidence in ontological knowledge bases}. In \bibinfo{booktitle}{\emph{WWW}}.
  ACM.
\newblock


\bibitem[\protect\citeauthoryear{Jain and Beetz}{Jain and Beetz}{2010}]%
        {jain2010soft}
\bibfield{author}{\bibinfo{person}{Dominik Jain} {and} \bibinfo{person}{Michael
  Beetz}.} \bibinfo{year}{2010}\natexlab{}.
\newblock \showarticletitle{Soft evidential update via markov chain monte carlo
  inference}. In \bibinfo{booktitle}{\emph{Annual Conference on Artificial
  Intelligence}}. Springer.
\newblock


\bibitem[\protect\citeauthoryear{Kadlec, Bajgar, and Kleindienst}{Kadlec
  et~al\mbox{.}}{2017}]%
        {kadlec2017knowledge}
\bibfield{author}{\bibinfo{person}{Rudolf Kadlec}, \bibinfo{person}{Ondrej
  Bajgar}, {and} \bibinfo{person}{Jan Kleindienst}.}
  \bibinfo{year}{2017}\natexlab{}.
\newblock \showarticletitle{Knowledge Base Completion: Baselines Strike Back}.
\newblock \bibinfo{journal}{\emph{arXiv preprint arXiv:1705.10744}}
  (\bibinfo{year}{2017}).
\newblock


\bibitem[\protect\citeauthoryear{Kok, Sumner, Richardson, Singla, Poon, Lowd,
  Wang, and Domingos}{Kok et~al\mbox{.}}{2009}]%
        {kok2009alchemy}
\bibfield{author}{\bibinfo{person}{Stanley Kok}, \bibinfo{person}{Marc Sumner},
  \bibinfo{person}{Matthew Richardson}, \bibinfo{person}{Parag Singla},
  \bibinfo{person}{Hoifung Poon}, \bibinfo{person}{Daniel Lowd},
  \bibinfo{person}{Jue Wang}, {and} \bibinfo{person}{Pedro Domingos}.}
  \bibinfo{year}{2009}\natexlab{}.
\newblock \bibinfo{booktitle}{\emph{The {A}lchemy system for statistical
  relational $\{$AI$\}$}}.
\newblock \bibinfo{type}{{T}echnical {R}eport}.
  \bibinfo{institution}{Department of Computer Science and Engineering,
  University of Washington}.
\newblock


\bibitem[\protect\citeauthoryear{Lehmann, Bizer, Kobilarov, Auer, Becker,
  Cyganiak, and Hellmann}{Lehmann et~al\mbox{.}}{2009}]%
        {dbpedia_jws_09}
\bibfield{author}{\bibinfo{person}{Jens Lehmann}, \bibinfo{person}{Chris
  Bizer}, \bibinfo{person}{Georgi Kobilarov}, \bibinfo{person}{S{\"o}ren Auer},
  \bibinfo{person}{Christian Becker}, \bibinfo{person}{Richard Cyganiak}, {and}
  \bibinfo{person}{Sebastian Hellmann}.} \bibinfo{year}{2009}\natexlab{}.
\newblock \showarticletitle{{DB}pedia - A Crystallization Point for the Web of
  Data}.
\newblock \bibinfo{journal}{\emph{Journal of Web Semantics}}
  \bibinfo{volume}{7}, \bibinfo{number}{3} (\bibinfo{year}{2009}),
  \bibinfo{pages}{154--165}.
\newblock
\urldef\tempurl%
\url{https://doi.org/doi:10.1016/j.websem.2009.07.002}
\showDOI{\tempurl}


\bibitem[\protect\citeauthoryear{Leit{\~a}o, Calado, and Weis}{Leit{\~a}o
  et~al\mbox{.}}{2007}]%
        {leitao2007structure}
\bibfield{author}{\bibinfo{person}{Lu{\'\i}s Leit{\~a}o},
  \bibinfo{person}{P{\'a}vel Calado}, {and} \bibinfo{person}{Melanie Weis}.}
  \bibinfo{year}{2007}\natexlab{}.
\newblock \showarticletitle{Structure-based inference of XML similarity for
  fuzzy duplicate detection}. In \bibinfo{booktitle}{\emph{CIKM}}. ACM.
\newblock


\bibitem[\protect\citeauthoryear{Liu, Wu, and Yang}{Liu et~al\mbox{.}}{2017}]%
        {liu2017analogical}
\bibfield{author}{\bibinfo{person}{Hanxiao Liu}, \bibinfo{person}{Yuexin Wu},
  {and} \bibinfo{person}{Yiming Yang}.} \bibinfo{year}{2017}\natexlab{}.
\newblock \showarticletitle{Analogical Inference for Multi-Relational
  Embeddings}.
\newblock \bibinfo{journal}{\emph{arXiv preprint arXiv:1705.02426}}
  (\bibinfo{year}{2017}).
\newblock


\bibitem[\protect\citeauthoryear{Macskassy and Provost}{Macskassy and
  Provost}{2005}]%
        {macskassy2005netkit}
\bibfield{author}{\bibinfo{person}{Sofus~A Macskassy} {and}
  \bibinfo{person}{Foster Provost}.} \bibinfo{year}{2005}\natexlab{}.
\newblock \showarticletitle{Netkit-srl: A toolkit for network learning and
  inference}. In \bibinfo{booktitle}{\emph{Proceeding of the NAACSOS
  Conference}}.
\newblock


\bibitem[\protect\citeauthoryear{Muggleton}{Muggleton}{1995}]%
        {muggleton1995inverse}
\bibfield{author}{\bibinfo{person}{Stephen Muggleton}.}
  \bibinfo{year}{1995}\natexlab{}.
\newblock \showarticletitle{Inverse entailment and Progol}.
\newblock \bibinfo{journal}{\emph{New generation computing}}
  \bibinfo{volume}{13}, \bibinfo{number}{3} (\bibinfo{year}{1995}),
  \bibinfo{pages}{245--286}.
\newblock


\bibitem[\protect\citeauthoryear{Nickel, Rosasco, and Poggio}{Nickel
  et~al\mbox{.}}{2015}]%
        {nickel2015holographic}
\bibfield{author}{\bibinfo{person}{Maximilian Nickel}, \bibinfo{person}{Lorenzo
  Rosasco}, {and} \bibinfo{person}{Tomaso Poggio}.}
  \bibinfo{year}{2015}\natexlab{}.
\newblock \showarticletitle{Holographic Embeddings of Knowledge Graphs}.
\newblock \bibinfo{journal}{\emph{AAAI}} (\bibinfo{year}{2015}).
\newblock


\bibitem[\protect\citeauthoryear{Niu, R{\'e}, Doan, and Shavlik}{Niu
  et~al\mbox{.}}{2011}]%
        {niu2011tuffy}
\bibfield{author}{\bibinfo{person}{Feng Niu}, \bibinfo{person}{Christopher
  R{\'e}}, \bibinfo{person}{AnHai Doan}, {and} \bibinfo{person}{Jude Shavlik}.}
  \bibinfo{year}{2011}\natexlab{}.
\newblock \showarticletitle{Tuffy: Scaling up statistical inference in markov
  logic networks using an rdbms}.
\newblock \bibinfo{journal}{\emph{VLDB}} \bibinfo{volume}{4},
  \bibinfo{number}{6} (\bibinfo{year}{2011}).
\newblock


\bibitem[\protect\citeauthoryear{Noessner, Niepert, and
  Stuckenschmidt}{Noessner et~al\mbox{.}}{2013}]%
        {noessner2013rockit}
\bibfield{author}{\bibinfo{person}{Jan Noessner}, \bibinfo{person}{Mathias
  Niepert}, {and} \bibinfo{person}{Heiner Stuckenschmidt}.}
  \bibinfo{year}{2013}\natexlab{}.
\newblock \showarticletitle{Rockit: Exploiting parallelism and symmetry for map
  inference in statistical relational models}.
\newblock \bibinfo{journal}{\emph{AAAI}} (\bibinfo{year}{2013}).
\newblock


\bibitem[\protect\citeauthoryear{P{\'e}rez, Arenas, and Gutierrez}{P{\'e}rez
  et~al\mbox{.}}{2009}]%
        {perez2009semantics}
\bibfield{author}{\bibinfo{person}{Jorge P{\'e}rez}, \bibinfo{person}{Marcelo
  Arenas}, {and} \bibinfo{person}{Claudio Gutierrez}.}
  \bibinfo{year}{2009}\natexlab{}.
\newblock \showarticletitle{Semantics and complexity of SPARQL}.
\newblock \bibinfo{journal}{\emph{ACM Transactions on Database Systems (TODS)}}
  \bibinfo{volume}{34}, \bibinfo{number}{3} (\bibinfo{year}{2009}),
  \bibinfo{pages}{16}.
\newblock


\bibitem[\protect\citeauthoryear{Riedel}{Riedel}{2008}]%
        {riedel08improving}
\bibfield{author}{\bibinfo{person}{Sebastian Riedel}.}
  \bibinfo{year}{2008}\natexlab{}.
\newblock \showarticletitle{Improving the accuracy and Efficiency of MAP
  Inference for Markov Logic}. In \bibinfo{booktitle}{\emph{Proceedings of the
  24th Annual Conference on Uncertainty in AI (UAI '08)}}.
  \bibinfo{pages}{468--475}.
\newblock


\bibitem[\protect\citeauthoryear{Schoenmackers, Etzioni, Weld, and
  Davis}{Schoenmackers et~al\mbox{.}}{2010}]%
        {schoenmackers2010learning}
\bibfield{author}{\bibinfo{person}{Stefan Schoenmackers}, \bibinfo{person}{Oren
  Etzioni}, \bibinfo{person}{Daniel~S Weld}, {and} \bibinfo{person}{Jesse
  Davis}.} \bibinfo{year}{2010}\natexlab{}.
\newblock \showarticletitle{Learning first-order horn clauses from web text}.
  In \bibinfo{booktitle}{\emph{EMNLP}}.
\newblock


\bibitem[\protect\citeauthoryear{Shekarpour, Marx, {Ngonga Ngomo}, and
  Auer}{Shekarpour et~al\mbox{.}}{2014}]%
        {SHE+14}
\bibfield{author}{\bibinfo{person}{Saeedeh Shekarpour}, \bibinfo{person}{Edgard
  Marx}, \bibinfo{person}{Axel-Cyrille {Ngonga Ngomo}}, {and}
  \bibinfo{person}{S{\"o}ren Auer}.} \bibinfo{year}{2014}\natexlab{}.
\newblock \showarticletitle{{SINA}: {S}emantic {I}nterpretation of {U}ser
  {Q}ueries for {Q}uestion {A}nswering on {I}nterlinked {D}ata}.
\newblock \bibinfo{journal}{\emph{Web Semantics}}  \bibinfo{volume}{1}
  (\bibinfo{year}{2014}), \bibinfo{pages}{--}.
\newblock


\bibitem[\protect\citeauthoryear{Stadler, Lehmann, H{\"o}ffner, and
  Auer}{Stadler et~al\mbox{.}}{2012}]%
        {SLHA11}
\bibfield{author}{\bibinfo{person}{Claus Stadler}, \bibinfo{person}{Jens
  Lehmann}, \bibinfo{person}{Konrad H{\"o}ffner}, {and}
  \bibinfo{person}{S{\"o}ren Auer}.} \bibinfo{year}{2012}\natexlab{}.
\newblock \showarticletitle{LinkedGeoData: A Core for a Web of Spatial Open
  Data}.
\newblock \bibinfo{journal}{\emph{Semantic Web Journal}} \bibinfo{volume}{3},
  \bibinfo{number}{4} (\bibinfo{year}{2012}), \bibinfo{pages}{333--354}.
\newblock
\urldef\tempurl%
\url{http://jens-lehmann.org/files/2012/linkedgeodata2.pdf}
\showURL{%
\tempurl}


\bibitem[\protect\citeauthoryear{Trouillon, Welbl, Riedel, Gaussier, and
  Bouchard}{Trouillon et~al\mbox{.}}{2016}]%
        {trouillon2016complex}
\bibfield{author}{\bibinfo{person}{Th{\'e}o Trouillon},
  \bibinfo{person}{Johannes Welbl}, \bibinfo{person}{Sebastian Riedel},
  \bibinfo{person}{{\'E}ric Gaussier}, {and} \bibinfo{person}{Guillaume
  Bouchard}.} \bibinfo{year}{2016}\natexlab{}.
\newblock \showarticletitle{Complex embeddings for simple link prediction}. In
  \bibinfo{booktitle}{\emph{International Conference on Machine Learning}}.
  \bibinfo{pages}{2071--2080}.
\newblock


\bibitem[\protect\citeauthoryear{Wuthrich}{Wuthrich}{1995}]%
        {wuthrich1995probabilistic}
\bibfield{author}{\bibinfo{person}{Beat Wuthrich}.}
  \bibinfo{year}{1995}\natexlab{}.
\newblock \showarticletitle{Probabilistic knowledge bases}.
\newblock \bibinfo{journal}{\emph{IEEE Transactions on Knowledge and Data
  Engineering}} \bibinfo{volume}{7}, \bibinfo{number}{5}
  (\bibinfo{year}{1995}), \bibinfo{pages}{691--698}.
\newblock


\end{thebibliography}
